# Magnetic characterization of sintered $MgB_2$ samples: effect of the substitution or "doping" with Li, Al and Si.

M. R. Cimberle, M. Novak[#], P. Manfrinetti*, A. Palenzona*

*CNR/INFM and Dipartimento di Fisica, via Dodecaneso 33, 16146 Genova, Italy*

[#] *Institute of Physics, V Holesovickach, 18000 Prague 8, Czech Republic*

- *INFM and Dipartimento di Chimica, via Dodecaneso 31, 16146 Genova, Italy*

*E-mail: cimberle@fisica.unige.it*

ABSTRACT

Powdered and sintered $MgB_2$ samples have been characterized through magnetic measurements performed from T = 5 K up to few degrees above the transition temperature of about 39 K. We found that the sintered samples behave as well-connected bodies, showing no trace of granularity. In order to obtain the critical current density value $J_c$ the "Critical State Model" has been therefore employed in a straightforward way. With the aim either to decrease the electron mean free path or to increase its $J_c$ we have attempted to introduce defects in the $MgB_2$ structure by different procedures: substitution of Lithium on the Magnesium site and doping of the precursor Boron powders with Aluminum and Silicon. The best result in terms of $J_c$ has been achieved by Silicon doping that, moreover, does not significantly affect the transition temperature.

## 1. INTRODUCTION

The recent discovery of the $MgB_2$ phase as an intermetallic superconductor with an exceptionally high transition temperature [1] has made this new and relatively "easy to do" superconductor attractive not only from a theoretical point of view, but also for technological applications. Since then a very fast and extended characterization of its physical properties has been carried out all over the world [2-10]. For the coherence length $\xi_0$ a value of 5.2 nm has been reported [2]; such a long $\xi_0$ value eliminates problems with weak links, which are instead typically present in "high-$T_c$" materials [11] and suggests the possibility to achieve a remarkably high critical current density if proper technological treatments are adopted. Therefore, the computation of the critical current density $J_c$, its temperature and field dependence and the possibility to increase it by properly adding pinning centers constitute a crucial problem to solve. We have attempted to introduce defects in the $MgB_2$ structure with two main purposes: firstly to decrease the mean free path of the normal electrons decreasing then the coherence length $\xi$ and increasing the $H_{c2}$ values, secondly to introduce pinning centers in order to make the irreversibility line steeper. We prepared different $MgB_2$ samples, with Li substitution on Mg site as well by doping with Si and Al. The results obtained through the standard procedure and the variations therein induced by our modifications are presented and discussed.

## 2. SAMPLE PREPARATION

The compounds $MgB_2$ and $Mg_{0.9}Li_{0.1}B_2$ were prepared by direct synthesis from the pure elements: Mg (in form of fine turnings, 99.999 wt.% purity), Li (pieces worked and cut out under dried organic solvent, 99.8 wt.% purity) and crystalline B (325-mesh powder, 99.7 wt.% purity) were well mixed together and closed by arc welding under pure argon into outgassed Ta crucibles, which were then sealed in quartz ampoules under vacuum. The samples were slowly heated up to 950 °C and maintained at this temperature for 1 day.

With regard to both Al- and Si-doped samples (both Al and Si 99.999 wt.% pure), a pre-reaction of B mixed with 0.5 at.% of the dopant element was carried out in Ta tubes, closed by the same procedure as above, and heat treated at 1000 °C – 4 day. The resulting doped B has been subsequently used for the synthesis with Mg. These two latter samples will be hereafter called $MgB_2$(Al) and $MgB_2$(Si), respectively. The choice of Al and Si as dopant elements for the first doping attempts was suggested by the existence of a terminal solid solubility in B up to few at.%. All the final products were gray-blackish powders, only slightly sintered; no reaction toward the crucible wall has been observed. They have been characterized by X-ray diffraction using a Guinier camera (CuK$\alpha$ radiation and pure Si as an internal standard). Samples, in prismatic shape, have been then prepared by pressing the powders in a stainless-steel die into a pellet, which was then sintered by heat treatment at 1000 °C - 2 days (again in Ta containers, welded under argon and closed in quartz tubes under vacuum). The density of the sintered samples is about 70% of the theoretical one; the sample final dimensions are approximately 2x2x3 mm.

Both the powders and sintered samples have been observed by SEM microscopy; SEM images have shown that the powders are constituted by agglomerates that show a wide size spread ranging from few μm up to clusters 150 μm large. Grains inside the sintered samples have smaller and more homogeneous size never exceeding 30 μm.

Magnetic measurements have been performed by a commercial SQUID magnetometer (Quantum Design).

The value of the lattice parameters of $MgB_2$, i.e.: $a$= 3.085(1) Å, $c$ = 3.524(1) Å, are in very good agreement with literature [12]; those of $Mg_{0.9}Li_{0.1}B_2$ and of both doped samples remained practically unchanged (both the $a$ and $c$ values within ±0.001 Å with respect to that of $MgB_2$). However, while X-ray diffraction patterns of Li-substituted sample and of the doped ones have not shown significant differences neither in the peaks positions nor in extra peaks, in the case of $Mg_{0.9}Li_{0.1}B_2$ sample a very slight variation in the intensity of some reflections was observed; a calculated X-ray pattern, by means of the Lazy-Pulverix program [13], agreed with the experimental one, confirming the effective substitution on Mg site by Li.

## 3. EXPERIMENTAL RESULTS AND DISCUSSION

In Fig.1 susceptibility measurements versus temperature at a constant field of 10 Gauss are shown for different samples: pure powders, pure sintered sample, powders from a reground sintered sample, Al-doped, Si-doped and Li-substituted sintered samples. All the measurements have been normalized to the constant value $\chi$ = -1 at low temperatures. The highest transition temperature is observed in the powders and in the pure sintered sample ($T_{c\ onset}$ = 38.8 K). As may be seen, substitution and doping change the transition temperature, indicating that the new element effectively entered in the structure. The strongest reduction in $T_c$ is produced by Al ($T_{c\ onset}$ = 37.2 K), in accordance with data in literature [14], while Li and Si produce smaller $T_c$ variations (less than 0.6 K and 0.4 K respectively). The transition widths are of about 1degree, the steepest transition being observed in the Si-substituted sample ($\Delta T_c$ = 0.75 K). Although the powders and the pure sintered sample show the same onset of the transition, in the latter the transition is worsened ($\chi$ = -0.5 is reached at T = 37.7 K in the powders and at T = 37.3 K in the sintered sample). We point out that this happens in spite of the potential counter-effect of the penetration depth that can decrease the shielded volume due to the small grain size.

In Fig.2 the irreversibility line (IL) and $H_{c2}$ for both powders and sintered samples are presented, together with IL for YBCO and $Nb_3Sn$ reference samples. In the x-axis the reduced temperature is reported. The IL has been measured by observing the deformation of the response function in the SQUID magnetometer while changing from an irreversible to a reversible behavior, as suggested in [15]. By this procedure, the IL for $J_c \rightarrow 0$ and therefore the highest $H^{g}_{irr}$ have been measured (the terminology is that of [16]). The upper critical field $H_{c2}$ has been identified by the onset of the superconducting transition (1% of the normal state value) measured at various magnetic fields. No significant variation in the $H_{c2}$ behavior has been observed neither in substituted or doped samples. They all exhibit at low fields the upward concavity as observed in the "clean limit", indicating that the introduction of disorder has not sufficiently reduced the normal electron mean free path. The IL of $MgB_2$ samples exhibits at low fields an upward curvature followed, at higher fields, by a linear behavior. What it is clear at glance is that, for all the samples, the IL lies well below that of YBCO and, at a greater extent, below that of the $Nb_3Sn$ sample. In particular what is distinctly different is the stronger field decrease of $J_c$ (see also the comparison between critical current densities in $MgB_2$ and $Nb_3Sn$ sintered samples reported in [2]), which makes clear the need of introducing effective pinning centers in $MgB_2$ samples.

The IL are similar for all the $MgB_2$ samples, with the exception of the Al substituted one, which stays distinctly lower. This trend could suggest a homogeneous behavior in term of $J_c$, but this is not true: at low temperature (T = 5 K), different amplitudes of the hysteresis loops indicate different values for the critical current density observed in the various samples. By measuring the hysteresis loops at T = 25 K we have verified that the closing down of the loop happens at lower fields for the Al doped sample, confirming then the IL trend and suggesting a different field dependence for Al-doped $MgB_2$ sample.

In Fig.3 the hysteresis loops measured at T = 5 K for all the sintered samples are shown together with the hysteresis loops measured both on the powders used to prepare the pure sintered sample and on powders obtained by regrinding a pure sintered sample. This has been done to check if the sintering process could change the intrinsic properties of the $MgB_2$ grains. All the samples show a similar decrease of the hysteresis loops by increasing the field: about one order of magnitude from zero up to 2 Tesla. The substitutions have not changed this general trend: their effect is essentially to enlarge the amplitude of the loops with respect to that observed in the pure sintered sample. At low fields (and for temperatures up to 15 K) many jumps are observed at the highest magnetization values. This is an indication of connected samples in which, due to the high value of $J_c$ and to the sample dimensions, a strong field gradient is established. Also in the starting powders a jump has been observed. We remember that powders exhibit a wide spread in the granulometry with clusters that can reach a dimension of 150 μm. The hysteresis loops on the parent powders are of the same order of magnitude of that on sintered samples. In the light of the great difference between the dimension of powders and sintered samples, this implies that powders present a very high $J_c$ with respect to the sintered samples.

Before calculating the critical current density from the hysteresis loops, it is necessary to assess if a sample behaves like a connected body or some level of granularity is present. To understand this, it is either possible to measure the remanent magnetization $M_{rem}$ as a function of the maximum magnetic field [17], or to check the shape of the hysteresis loop. In particular, if the sample is totally connected, the derivative of the virgin curve or of the reverse leg of the magnetization follows a well-defined analytical behavior. The results are well known in the case of a field constant $J_c$ and depend on the sample shape (slab, cylinder etc) [18], but if $J_c$ is strongly field dependent it is impossible to define an analytical solution. Therefore, we have examined the derivative of the reverse leg of the magnetization at high fields, where a constant $J_c$ approximation is better satisfied. This analysis allows moreover to determine the value of the full penetration field $\mu_0 H_p$ at a certain $\mu_0 H_{max}$ value, and therein the $J_c(H)$ value.

In Fig. 4 a) we show $M_{rem}$ versus $\mu_0 H_{max}$ as obtained on the pure sintered sample at T = 5 K. The shape of the curve, that presents a one step transition, clearly indicates absence of granularity. The same was observed at T = 25 K. As a further confirmation of a "totally connected body" behavior, in Fig.4 b) the derivative of the reverse leg of magnetisation as observed in the same pure sample at T = 25 K and $\mu_0 H_{max}$ = 0.6 T is shown. A fit of the experimental data with the relationship $\frac{dM}{dH} = -1 + \frac{2 \cdot (H_{max} - H)}{3H_p} - \frac{(H_{max} - H)^2}{(3H_p)^2}$ (1), valid for a bulk superconductor with a field constant $J_c$ [18], is reported as well. The reliability of the fit confirms the absence of granularity. The $J_c$ calculation may be therefore performed by the classical formulas of the "Critical State Model". In particular, to take into account the parallelepiped shape of our samples, we have modeled the magnetization profile inside the sample like an "inverted roof" [19] with the same slope in all the directions (that means no anisotropy of the critical current density). In table 1, the $J_c$ values calculated for the various samples at T = 5 K and T = 25 K are shown. We point out that in any case $J_c$ increases by increasing the doping, ranging from $5\times10^4$ A/cm$^2$ for the pure sintered samples up to $1.5\times10^5$ A/cm$^2$ for the Si doped sample at T = 5 K and zero field. Going to $\mu_0 H_{ext}$ = 2 T, $J_c$ decreases of about a factor 10. In table 1 the $J_c$ values relative both for the parent powders and for the reground powders are reported. Because of the large spread in the powder sizes, to estimate $J_c$ the highest $\mu_0 H_p$ value, as obtained from the derivative of the reverse leg of magnetization, has been attributed to the largest observed cluster dimension (150 μm for parent powders and 70 μm for the reground ones). The obtained values are much higher than for sintered samples, but the same strong field dependence is observed. This is consistent with the estimation of the activation energy $U_0$ obtained by magnetization decay measurements performed at $\mu_0 H_{ext}$ = 0.2 T and 2 T at T = 10 K and analyzed with the standard Anderson creep theory [20]. The $U_0$ values also decrease of about one order of magnitude: we found out 0.8 and 0.067 eV respectively. A decrease of a factor 2 in $J_c$ in the reground powders may be attributed to a variation of intrinsic properties of $MgB_2$ grains that have been pressed, sintered and reground.

In the light of the presented data the decrease in $J_c$ of about one order of magnitude observed in the sintered sample with respect to the powder does not appear as a "weak link effect", as in high $T_c$ superconductors. The grain boundaries have the same superconducting properties of the grains, and the sintered samples behave, from the magnetic point of view, like well-connected bodies. It appears reasonable due to the small surface contact between the grains that, for their hardness, are difficult to compact. Moreover, the sintering procedure can change the intrinsic superconducting properties of the material.

Finally, the temperature variation of the critical current density has been measured for Al doped samples. The results are presented in Fig. 5 as M (H = 0 T, 0.6 and 1 T) versus temperature. A linear decrease is observed both in the remanent magnetization and in the higher field magnetization.

## 4. CONCLUSIONS

Powdered and sintered samples of $MgB_2$ pure and doped with Li, Si and Al have been characterized by a set of magnetic measurements. Neither $H_{c2}$ nor the irreversibility line have shown a significant change in the temperature and field range we have explored. Nevertheless, hysteresis loops measured at low temperature indicate an improvement of the critical current density for all the chemical treatments, the best result being achieved by the Si doping (about a factor 3 in the $J_c$ values). So, also if the mean free path is not significantly reduced, Si or Al act as pinning centers at least at low temperature, suggesting their presence as clusters. The IL analysis makes clear that stronger pinning centers must be introduced to make this material more appealing for technological applications. This may be assessed by comparing the IL with that typical of low $T_c$ superconductors (see Fig.2). In particular, the field dependence of the critical current density appears to be very strong, while, by increasing the temperature from 5 K up to 30 K the critical current decreases only by a factor 6. No trace of magnetic granularity has been found in our sintered samples. Therefore it is reasonable to attribute the strong decrease of the critical current density of the sintered samples with respect to the powder to the low density of the produced samples (small connection area between grains) and also to intrinsic variations due to the sintering procedure (for example relaxation of defects during the 2 days long heat treatment). Anyway, the lack of "weak links" suggests that potentially high $J_c$ values are obtainable: with this purpose, both the sample density and the pinning strength must be increased.

## ACKNOWLEDGEMENTS

Two of us (A.P. and P.M.) thank the Italian *Ministero della Ricerca Scientifica e Tecnologica* (*M.U.R.S.T.*) for financial support, this work being part of the National Research Program "Alloys and Intermetallic Compounds: thermodynamics, physical properties, reactivity". M.N. is thankful to ESF Vortex Program.

REFERENCES

[1] Nagamatsu J., Nakagawa N., Muranaka T., Zenitani Y., Akimitzu J., 2001 *Nature* **410** 63-64
[2] Finnemore D.K., Ostenson J.E., Bud'ko S.L., Lapertot G., Canfield P.C., 2001 Cond-mat/0102114
[3] Bugoslavsky Y., Perkins G.K., Cohen L.F., Caplin A.D., 2001 Submitted to *Nature*
[4] Kambara M., Hari Babu N., Sadki E.S., Cooper J.R., Minami H., Cardwell D.A., Campbell A.M., Inoue I.H., 2001 *Supercond. Science and Technol.* **14** L5-L7
[5] Lorentz B., Meng L., Xue Y.Y., Chu C.W., 2001 cond-mat/0104041
[6] Kang W.N., Jung C.U., Kijoon H.P., Park Min-Seok, Lee S.Y., Kim Hyeong-Jin, Choi Eun-Mi, Kim Kyung Hee, Kim Mun-Seog, Lee Sung-Ik, 2001 cond-mat/0102313
[7] Larbalestier D.C., Cooley L.D., Rikel M.O., Polyanskii A.A., Jiang J., Patnaik S., Cai X.Y., Feldmann D.M., Gurevich A., Squitieri A.A., Naus M.T., Eom C.B., Hellstrom E.E., Cava R.J., Regan K.A., Rogado N., Hayward M.A., He T., Slusky J.S., Khalifah P., Inumaru K., Haas M., 2001 *Nature* **410** 186-189
[8] Dhalle' M., Toulemonde P., Beneduce C., Musolino N., Decroux M., 2001 submitted to *Physica C*
[9] Bud'ko S.L., Petrovic C., Lapertot G., Cunningham C.E., Canfield P.C., 2001 cond-mat/0102413
[10] Fuchs G., Müller K.-H., Handstein A., Nenkov K., Narozhnyi V.N., Eckert D., Wolf M., Schultz L. 2001 cond-mat/0104088
[11] Cimberle M. R., Ferdeghini C., Flükiger R., Giannini E., Grasso G., Marré D., Putti M., Siri A.S., 1995 *Physica C* **251** 61-67
[12] Naslain MM.R., Guette A., Barret M., 1973 *J. Solid State Chem.* **8** 68-85
[13] Yvon K., Jeitschko W., Parthé E., 1977 *J. Appl. Crystallogr.* **10** 73-76
[14] Slusky J.S., Rogado N., Regan K.A., Hayward M.A., Khalifah P., He T., Inumaru K., Loureiro S., Haas M.K., Zandbergen H.W., Cava R.J. 2001, cond-mat/0102262
[15] Suenaga M., Welch D.O., Budhani R., 1992 *Supercond. Sci. Technol.* **5** S1-S8
[16] Wen H.H., Li S.L., Zhao Z.W., Ni Y.M., Ren Z.A., Che G.C., Zhao Z.X., 2001 cond-mat/0103521
[17] Müller K.-H., Andrikidis C., Liu H.K., Dou S.X. 1994 *Phys. Rev B* **50** 10218-10224
[18] Bean C.P. 1964 *Rev. Modern Phys*…..? 31-39
[19] Gyorgy E.M., van Dover R.B., Jackson K.A., Schneemeyer L.F., Waszczk J.V. 1989 *Appl. Phys. Lett.* **55** 283
[20] Anderson P.W. 1962 *Phys. Rev. Lett.* **9** 309

FIGURE CAPTIONS:

Fig.1: Susceptibility versus temperature for powdered and sintered samples in an applied magnetic field of 10 Gauss.

Fig.2: Irreversibility Line and $H_{c2}$ for powdered and sintered samples. For a comparison the IL for YBCO and $Nb_3Sn$ samples are shown. IL is indicated by continuous lines, $H_{c2}$ by dotted lines. The same symbol refers to the same sample.

Fig.3: Hysteresis loops measured at T =5 K for the various samples as indicated in the legend.

Fig.4 a): $M_{rem}$ versus $\mu_0 H_{max}$ at T = 5 K for the pure sintered $MgB_2$ sample.

Fig.4 b): Derivative of the reverse leg of the magnetization dM/dH versus $\mu_0$ ($H_{max}$ –H) as measured at $\mu_0 H_{max}$ = 2 T and T = 25 K in the pure sintered sample. (Triangle: experimental data; Circles with continuous line: fit by the "Critical State Model" (see text)).

Fig.5: Magnetization versus temperature in Al-doped sample.

Table 1: $J_c$ values (in A/cm$^2$) valued by magnetic measurements at various fields and temperatures.

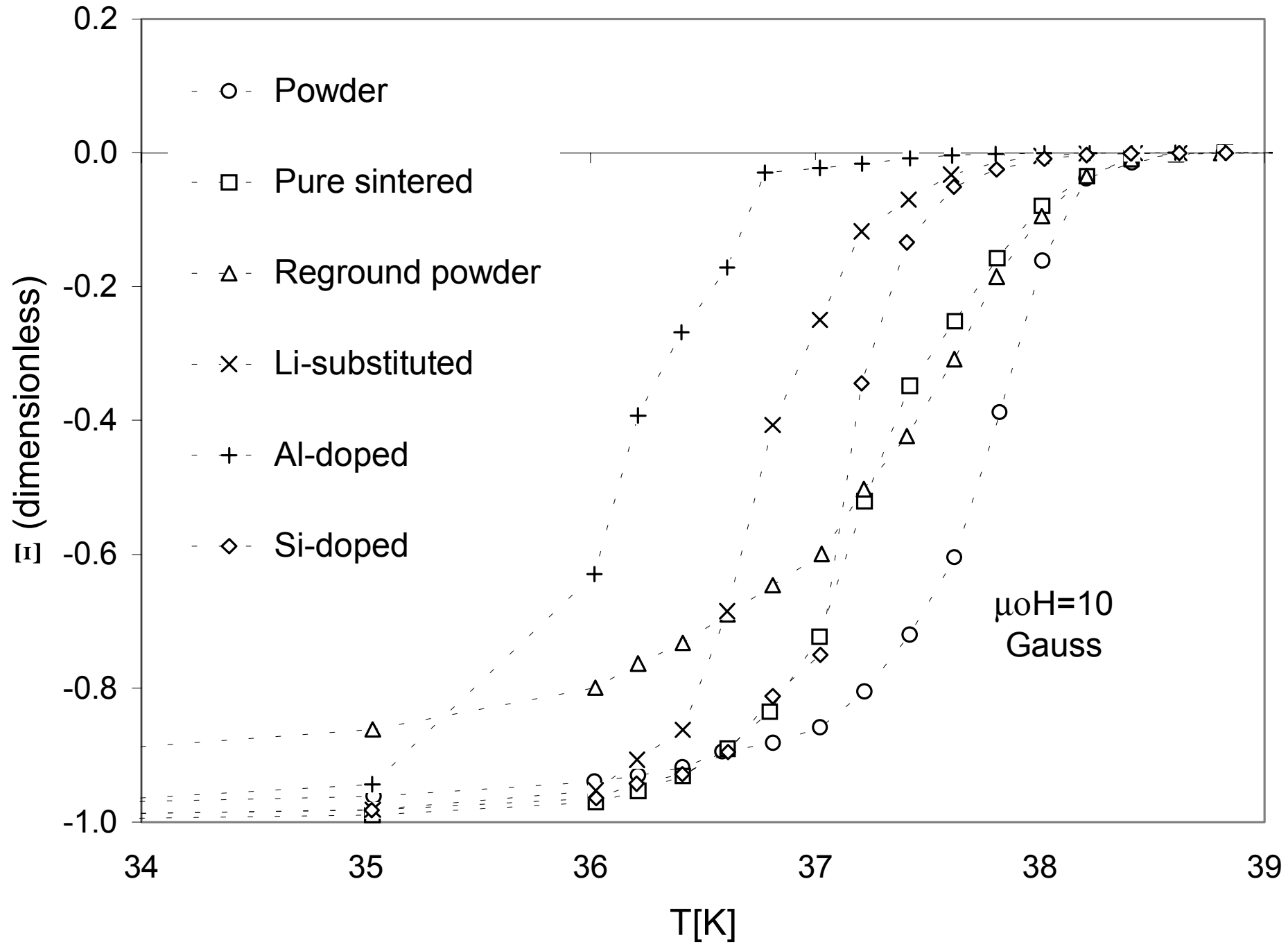

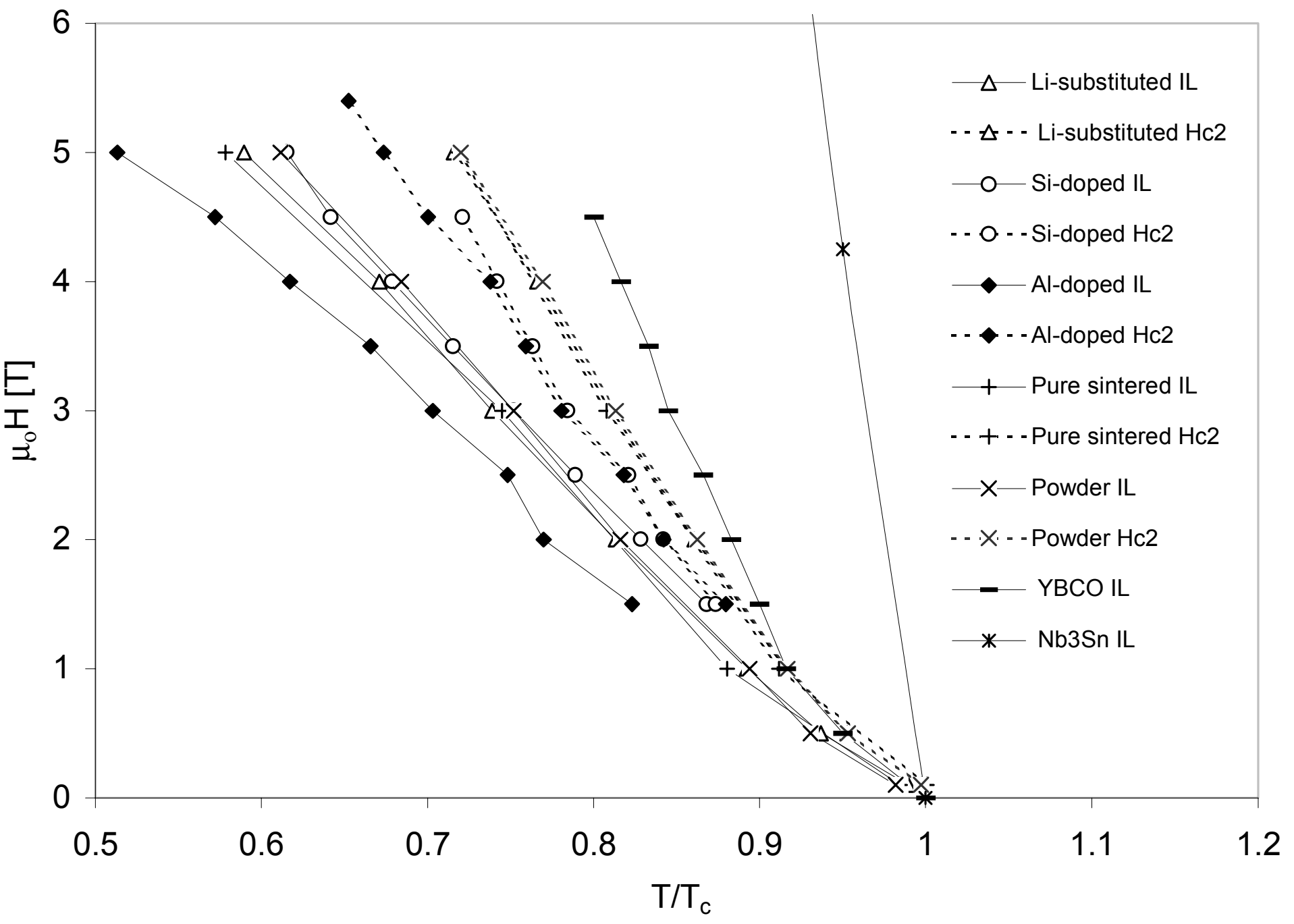

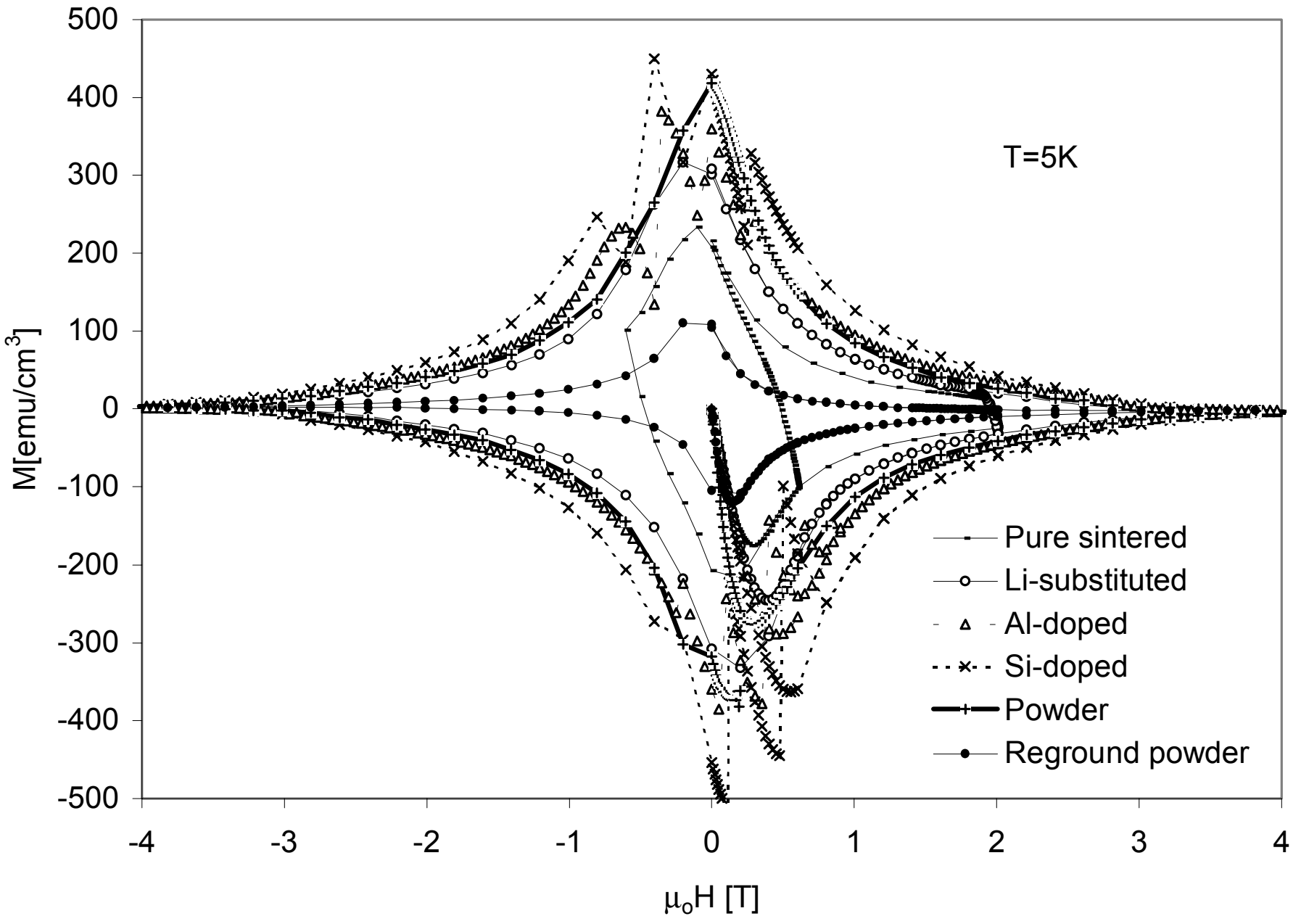

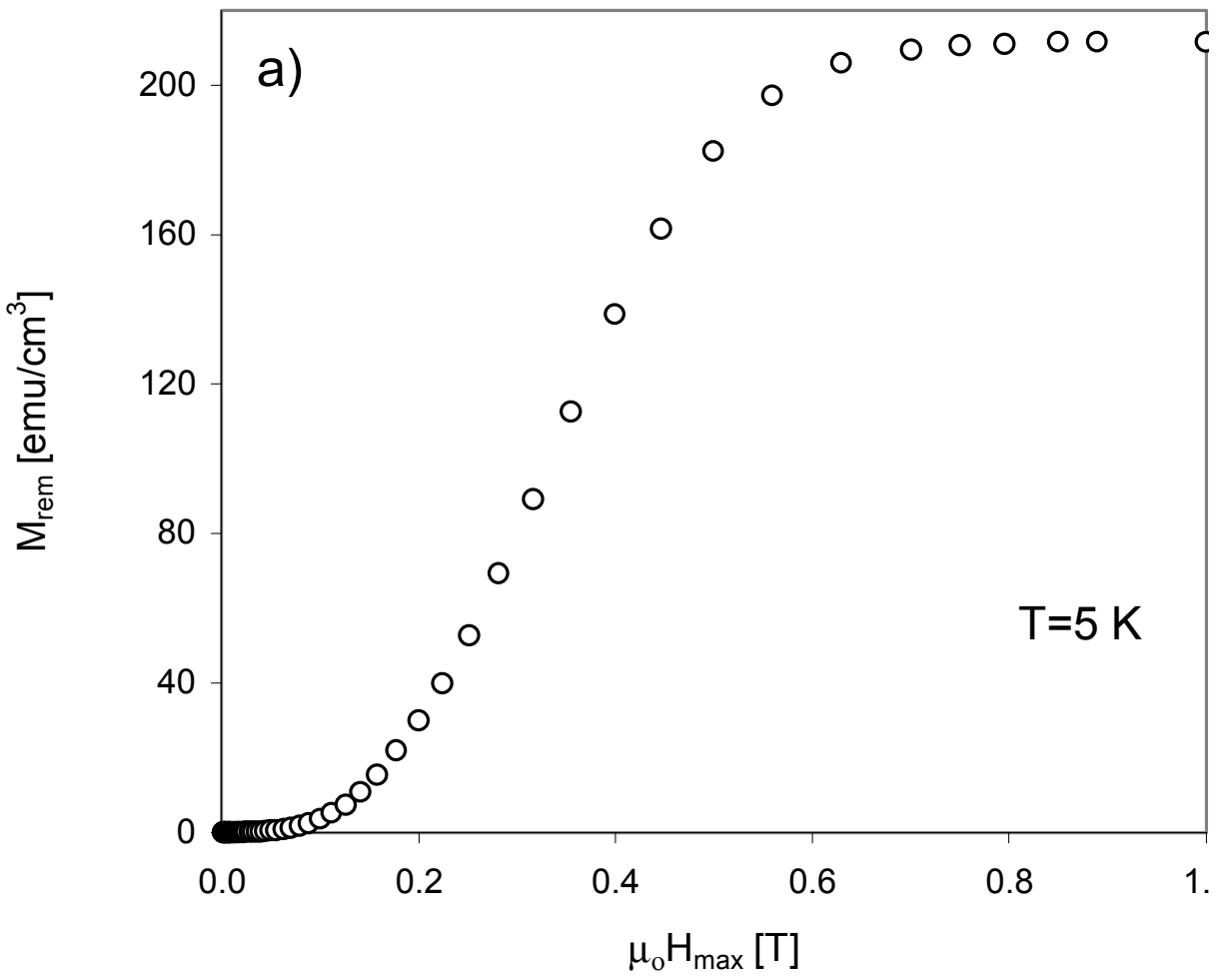
a)
T=5 K
Mrem [emu/cm3]
μoHmax [T]

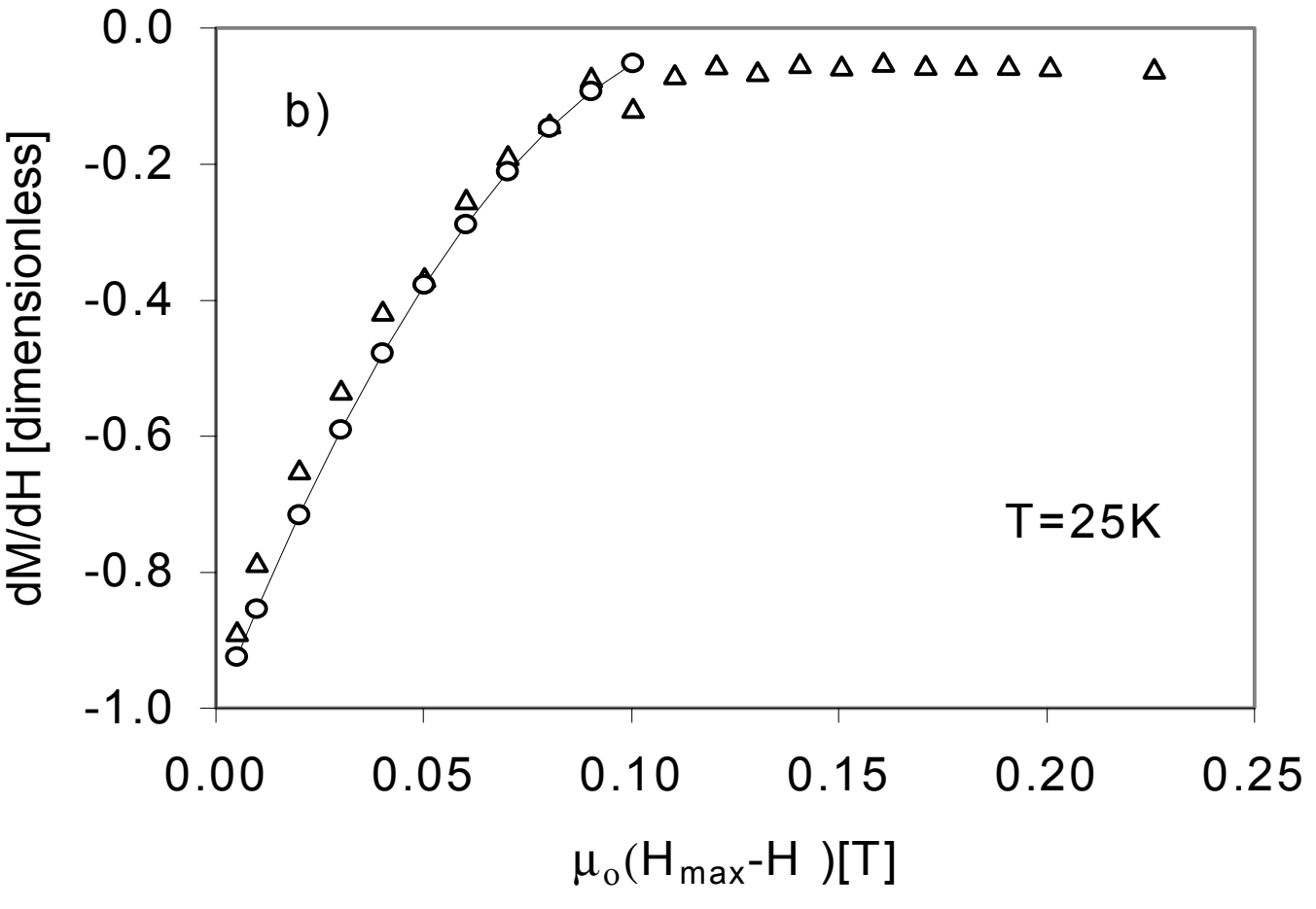
b)
T=25K
dM/dH [dimensionless]
μo(Hmax-H )[T]

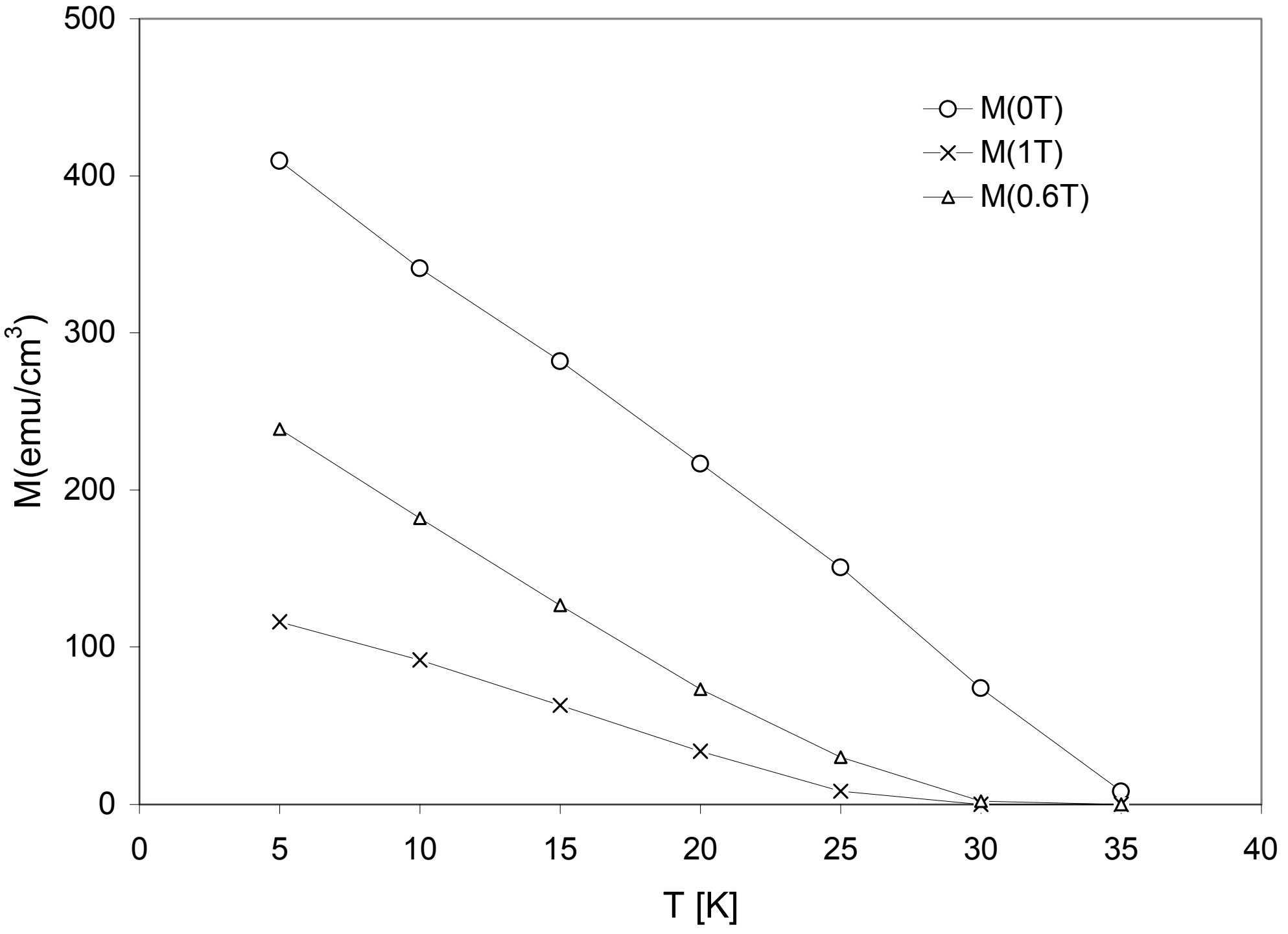

| | T=5K $\mu_0H$=0T | T=5K $\mu_0H$=2T | T=25K $\mu_0H$=0T | T=25K $\mu_0H$=0.6T |
|---|---|---|---|---|
| Pure sintered | $5\times10^4$ | $5\times10^3$ | $2\times10^4$ | $5\times10^3$ |
| Si-doped | $1.5\times10^5$ | $1.5\times10^4$ | $6\times10^4$ | $1.5\times10^4$ |
| Al-doped | $1\times10^5$ | $1\times10^4$ | $4\times10^4$ | $8\times10^3$ |
| Li substituted | $7\times10^4$ | $7\times10^3$ | | |
| Powder | $2\times10^6$ | $2\times10^5$ | | |
| Reground powder | $1\times10^6$ | $1\times10^5$ | | |